%% file: main.tex
\documentclass[conference]{IEEEtran}
\IEEEoverridecommandlockouts

\usepackage{cite}
\usepackage{amsmath,amssymb,amsfonts}
\usepackage{algorithmic}
\usepackage{graphicx}
\usepackage{textcomp}
\usepackage{xcolor}
\usepackage{algorithm}
\usepackage{latexsym}
\usepackage{hyperref}
\usepackage{url}
\input{macros}

\usepackage{bbm}
\usepackage{subcaption}
\usepackage{tabularx}
\usepackage{multirow}
\usepackage{verbatim}
\usepackage{color}
\usepackage{mathrsfs}
\usepackage{float}
\usepackage{bm}
\usepackage{booktabs}
\usepackage{diagbox}

\include{defn}
\def\BibTeX{{\rm B\kern-.05em{\sc i\kern-.025em b}\kern-.08em
    T\kern-.1667em\lower.7ex\hbox{E}\kern-.125emX}}

\begin{document}

\title{Localization of Coordinated Cyber-Physical Attacks in Power Grids Using Moving Target Defense and Deep Learning }

\author{\IEEEauthorblockN{\textbf{Yexiang Chen}\IEEEauthorrefmark{1}, \textbf{Subhash~Lakshminarayana}\IEEEauthorrefmark{1}, \textbf{Fei Teng}\IEEEauthorrefmark{2}}

\IEEEauthorblockA{\IEEEauthorrefmark{1}School of Engineering, University of Warwick, Coventry, UK}
\IEEEauthorblockA{\IEEEauthorrefmark{2}Department of Electrical and Electronic Engineering, Imperial College London, UK}}

\maketitle

\begin{abstract}
As one of the most sophisticated attacks against power grids, coordinated cyber-physical attacks (CCPAs) damage the power grid's physical infrastructure and use a simultaneous cyber attack to mask its effect. This work proposes a novel approach to detect such attacks and identify the location of the line outages (due to the physical attack). The proposed approach consists of three parts. Firstly, moving target defense (MTD) is applied to expose the physical attack by actively perturbing transmission line reactance via distributed flexible AC transmission system (D-FACTS) devices. MTD invalidates the attackers' knowledge required to mask their physical attack. Secondly, convolution neural networks (CNNs) are applied to localize line outage position from the compromised measurements. Finally, model agnostic meta-learning (MAML) is used to accelerate the training speed of CNN following the topology reconfigurations (due to MTD) and reduce the data/retraining time requirements. Simulations are carried out using IEEE test systems. The experimental results demonstrate that the proposed approach can effectively localize line outages in stealthy CCPAs.   
\end{abstract}


\section{Introduction} \label{sec:Introduction}
Integrating information and communication technologies (ICTs) increases the efficiency and reliability of modern power systems. However, the ICTs make power grid infrastructure vulnerable to malicious attacks. In particular, coordinated cyber-physical attacks (CCPAs) can be a major threat to the power grid \cite{Soltan2015, LiCCPA2016, DengCCPA2017}. It consists of a ``physical attack'' such as disconnecting a transmission line/circuit breaker, and a coordinated ``cyber attack'' such as a false data injection attack (FDIA) that masks the effect of the physical attack. Undetected CCPAs can have a major impact, such as triggering cascading failure across the grid. 

Existing work on defense against CCPAs mainly focuses on attack detection, i.e., detecting the existence of an attack. The first approach relies on installing known-secure phasor measurement units (PMUs) to ensure that the attacker cannot inject an undetectable CCPA \cite{LiCCPA2016, DengCCPA2017} or analyses transient data from PMUs assuming that only a limited number of them can be compromised by the attacker \cite{JamesPMU2021}. 
However, security updates such as PMU installation across the system could be expensive, and PMUs themselves can be vulnerable to attacks (e.g., spoofing on GPS receivers \cite{Daniel2012}). Alternatively, machine learning (ML) techniques have been used extensively for attack detection in power grids \cite{HeDL2017, ZhangML2020, Chenguang2020}. However, ML techniques can be vulnerable to adversarial examples \cite{SaygheISGT2020, Kamal2021}. As we will show in this work, a cleverly crafted coordinated FDIA that masks the effect of the physical attack can significantly degrade the performance of ML algorithms. Thus, the straightforward application of ML algorithms does not perform effectively in attack detection/localization. 

To overcome these drawbacks, the technique of moving target defense has been shown to be particularly effective \cite{LiuMTD2018, LakshMTD2021, ZhangMTD2020, Higgins2021,  Lakshminarayana2021}. The key idea is to actively introduce perturbations to the power grid's line reactances (using distributed flexible AC transmission -- DFACTS -- devices) to invalidate the attacker's knowledge, thereby limiting the time window for the attacker to learn the system settings and craft undetectable attacks. The majority of the work on MTD focuses on detecting FDI attacks. The main focus of these works lies in formulating metrics to design effective MTD \cite{LiuMTD2018, LakshMTD2021, ZhangMTD2020}, formulating hidden MTD whose activation cannot be detected by the attacker \cite{Higgins2021}, and exposing the cost-benefit trade-off \cite{LakshMTD2021}. In the specific context of CCPAs, recent work \cite{Lakshminarayana2021} formulated optimal MTD design. Specifically, they proposed optimal D-FACTS placement to detect CCPAs as well as a game-theoretic framework to limit the MTD cost. However, to the best of our knowledge, most of the existing works of MTD focus on the problem of attack detection only. In contrast, the problem of attack localization, i.e., identifying the disconnected lines in CCPA has not received attention. 

In the context of CCPAs, attack localization, specifically identifying the physically disconnected lines,  is challenging since it involves localization from compromised measurements. To overcome this research gap, we propose a novel approach by combining MTD and ML methods. 
The proposed approach consists of three parts. Firstly, MTD is applied to invalidate the cyber mask and expose the line outages. Secondly, ML models are developed to identify the location of line outages from the compromised measurements. Thirdly, model agnostic meta-learning (MAML) is utilized to retrain DNN and track the system's changing baseline (due to MTD perturbations), thus improving the ML model's training speed and reducing the data requirements. We summarize our main contributions in the following:

\begin{itemize}
    \item We analyze the mechanism of CCPAs under the DC power flow model and propose three kinds of CCPA named ``partial CCPA'', ``extra CCPA'' and ``full CCPA''. We demonstrate using IEEE test systems that the existing ML methods can only localize ``partial CCPA'' but can not localize sophisticated attacks such as ``extra CCPA'' and ``full CCPA''.    
   
    \item To defend against CCPAs, we develop an attack localization model by combining MTD with ML methods, which are used to expose stealthy attacks and identify the attack location, respectively.
   
    \item To make the proposed defense more practical, we improve the attack localization model using MAML, which accelerates the training speed and reduces the data requirement. By doing so, the developed model can adapt to the topology reconfigurations caused by MTD and planned system operations.
   
    \item We conduct extensive simulations using the benchmark IEEE bus system. The experimental results show that the proposed approach can effectively localize line outages in stealthy CCPAs.  
\end{itemize}

The rest of this paper is organized as follows. Section~\ref{sec:Preliminaries} introduces the power system model and CCPAs. Section~\ref{sec:CCPA identification} details the proposed approach to localize line outages in stealthy CCPAs and \ref{sec:Meta Learning} describes the MAML approach. Section~\ref{sec:Simulations Settings A} presents the simulation setting. Section~\ref{sec:Simulation Results} analyses the simulation results to show the effectiveness of the proposed approach. The conclusions are presented in Section~\ref{sec:Conclusions}. 

\addtolength{\topmargin}{0.04in} 

\section{Preliminaries} \label{sec:Preliminaries}

\subsection{Power Grid Model} \label{sec:Power Grid Model}
We consider a power grid consist of a set $\mathcal{N} = \{1,2,  \ldots,N\}$ of buses and a set $\mathcal{L} = \{1,2,  \ldots,L\}$ of transmission lines. The generation and load on bus $i$ are denoted as $G_i$ and $L_i$, respectively. The power injection on bus $i$ is given by $ P_i = G_i - L_i$. We let $l = \{ i, j\}, \ i \neq j$ denote a transmission line $l \in \mathcal{L}$ that connects bus $i$ and bus $j$, and its reactance by $x_l$. The voltage phase angle of bus $i$ is denoted by $\theta_i$. Considering the DC power flow model, we use $F_l = {1 \over x_l} (\theta_i - \theta_j)$ to denote the directed active power flow, which points from bus $i$ to bus $j$. 

\emph{Power System State Estimation:} The PSSE finds the best estimation of the system state from the noisy measurements. We consider the DC power flow model henceforth. The state of the system is given by the voltage phase angles $\thetav = (\theta_1, \theta_2, \ldots, \theta_N)^T$. We use $ \zv = (z_1, z_2, \ldots, z_m)$ to denote the available measurements, where $m$ is the number of meters, and $m \geq n$. The measurement error (noise) is denoted by $ \ev = (e_1, e_2, \ldots, e_m)$ which is assumed to be Gaussian. The relationship between measurements and state variables can be represented as:
\begin{align}
     \zv = \Hm \thetav + \ev,
\end{align}
where, in this case, the measurement $\zv \in \RR^m$ consists of active power injection, active power flow and reverse active power flow, i.e. ${\zv} = [\tilde{\Pm},\tilde{\Fm},-\tilde{\Fm}]^T$, where $\Pm = (P_1, P_2, \ldots, P_N)$, $\Fm = (F_1, F_2, \ldots, F_L)$. Let $\Am \in \RR^{(N-1) \times L}$ denote the reduced branch-bus incidence matrix obtained by removing the row of the slack bus and $\Dm \in \RR^{L \times L}$ as a diagonal matrix of the reciprocals of link reactances. Then, the system's Jacobian matrix $\Hm \in \RR^{m \times n}$ is given by $\Hm =  [\Dm \Am^T;-\Dm \Am^T;\Am \Dm \Am^T]$. Using the minimum mean squared estimation method, the estimate of the system state is given by,
\begin{align}
    \hat{\thetav} = (\Hm^T \Wm \Hm)^{-1} \Hm^T \Wm \zv,
\end{align}
where $\Wm = \diag(\sigma_1^{-2}, \sigma_2^{-2}, \ldots, \sigma_m^{-2})$ is a diagonal matrix, and $\sigma_i, i = 1, \ldots,m$ is the standard deviation of sensor measurement noise.

Bad data detection (BDD) is based on the measurement residual, which defined as $\rv =  \zv - \Hm \hat{\thetav} $. If the Euclidean norm of measurement residual exceeds a specific threshold $ {\lVert \rv \rVert}_2  \geq \tau$, the BDD will trigger the alarm to indicate the presence of bad data (e.g., faulty measurements and/or attack).

\emph{False Data injection Attack:} FDIA injects malicious data into the measurements, misleading PSSE to obtain an incorrect system state estimation. We denote an FDIA vector by $\av = (a_1, a_2, \ldots, a_m)^T$. Then the compromised measurement is given by $\zv_c = {\bf z} + {\bf a}$. If the attack vector follows the constraint $\av = \Hm \cv $, it will not change the measurement residual and thus will not be detected by BDD \cite{Liu2011}. The corresponding estimate of the system state is given by $\hat{\thetav}_c = \hat{\thetav} + \cv$, where $\cv = (c_1, c_2, \ldots, c_n)^T$ is the estimation error due to the attack. 

\subsection{Coordinated Cyber-Physical attack} \label{sec:Coordinated Cyber-Physical attack}
While the FDIA harms the power system by modifying sensor measurement and misleading system operators to take incorrect operations, CCPA damages the power grid physically and uses the coordinated FDIA to mask the effect of the attack \cite{DengCCPA2017}. In this research, we mainly consider the physical attack in form of a transmission line disconnection. Under DC power flow, the relationship between pre-attack and post-physical-attack measurements can be derived in a straightforward manner as
\begin{align}
    \zv_p &=  \zv + \Hm \Delta \bm{\theta} + \Delta \Hm \bm{\theta}_p,
\end{align}
where the subscript ``p'' represents the power grid parameters after the physical attack, $ \zv$ and $\zv_p$ denote pre-attack and post-physical-attack measurements, respectively. $\av_p \defines  \Hm \Delta \bm{\theta} + \Delta \Hm \bm{\theta}_p $ is the overall change in the measurement because of the physical attack. $\Delta \Hm $, and $\Delta \bm{\theta}$ are the change in the system's Jacobian matrix and phase angle before and after physical attack, given by $\Delta \Hm = \Hm_p - \Hm $ and $\Delta \bm{\theta} = \bm{\theta}_p - \bm{\theta} $. The change in the measurements due to the physical attack will increase the residual value, and the attack will likely be detected by the BDD. Under CCPA, the attacker injects a coordinated FDIA following the physical attack to avoid detection. The simplest FDIA that can mask the effect of the physical attack is of the form  $\av_c = -\Delta \Hm \bm{\theta}_p$, which leads measurement to  
\begin{align}
     \zv_{ccpa} = \zv_p + \av_c 
     = \zv + \Hm \Delta \bm{\theta}  \label{eqn:CCPA_P}.
\end{align}
Comparing with the pre-physical-attack measurements $\zv$, the only difference in the post-CCPA measurements $\zv_{ccpa} $ is the term $\Hm \Delta \bm{\theta}$. Such $\zv_{ccpa}$ will not cause an increase in the value of the residual and will be identified as a normal system measurement by BDD. 
We name the CCPA in \eqref{eqn:CCPA_P} as ``partial CCPA'', since it only hides a part of the change in measurements due to the physical attack (i.e., only removes the $ \Delta \Hm \bm{\theta}_p$  component from $\zv_p,$ while leaving the $\Hm \Delta \bm{\theta}$ component). To launch a ``partial CCPA'' (i.e., $\av_c = -\Delta \Hm \bm{\theta}_p$), attackers can calculate $ \Delta \Hm$ using the tripped branch reactance, and obtain $\bm{\theta}_p$ using the reactance of one alternative path between the two ends of the tripped branch. \cite{Lakshminarayana2021}.

Alternatively, the attacker can improve the stealthiness of CCPA to avoid detection from advanced detectors (e.g., ML based). Let us consider a CCPA with an extra FDIA:
\begin{align}
    \zv_{ccpa\_extra} 
    &= \zv_{ccpa} + \av_{extra}  \nonumber\\
    &= \zv + \Hm \Delta \bm{\theta} + \Hm \cv \nonumber\\ 
    &= \zv + \Hm (\Delta \bm{\theta} + \cv). \label{eqn:CCPA_E}
\end{align}
In this attack, the attacker injects an additional FDIA $\av_{extra} = {\bf H} \cv, \cv \in \mathbb{R}^N$, which is also undetectable to BDD. ${\bf c}$ is the estimation distortion and can be a random vector with the dimension of $N$. We name the CCPA in \eqref{eqn:CCPA_E} as ``extra CCPA'' since it injects an additional FDIA to disturb the detection. Note that an attack of this form can also be constructed with partial knowledge of the system topology and can be constructed in a sparse manner. 

Furthermore, the attackers can completely mask the effect of the line outage using a more sophisticated FDI attack. Specifically, consider an FDIA of the form $\av_{full} = \zv - \zv_p$. If $\av_{full}$ is injected following the physical attack, then we have, 
\begin{align}
    \zv_{ccpa\_full} &= \zv_p + \av_{full} \nonumber\\
    & = \zv_p + \zv - \zv_p = \zv. \label{eqn:CCPA_F}
\end{align}
Note that to compute $\av_{full}$, before launching the physical attack, the attacker must first compute the power flows following the line disconnection (e.g., using line outage distribution factors) to obtain the post-physical-attack measurement $\zv_p$. Note that ``full CCPA'' may potentially require complete knowledge of the power grid topology (e.g., with multiple line disconnections) and injecting an FDIA in all the measurements of the system. In contrast, the ``partial CCPA'' and ``extra CCPA'' are sparse vectors and hence, require less effort from the attacker (in terms of the ``write'' access on system measurements). 

\section{CCPA identification based on deep neural network and moving target defense}  \label{sec:CCPA identification}

In this section, we propose CCPA localization method by combining ML with MTD. We describe the proposed defense step by step in the following.

\subsection{DNN-based Line Outage Identification with Uncompromised Measurements} \label{sec:DNN Approaches for CCPA Identification}
We firstly review DNN-based line outage localization based on uncompromised measurements (i.e., without a coordinated cyber attack) \cite{Zhaoyue2020}. Specifically, we develop a DNN approach that can build a mapping between the measurements and the line that is under outage. Let us denote the DNN's input, ground truth labels, and the DNN's output as $\zv = (z_1, z_2, \ldots, z_m)$, $\yv = (y_1, y_2, \ldots, y_L)$, $\hat{\yv} = (\hat{y}_1, \hat{y}_2, \ldots, \hat{y}_L)$, respectively. Herein, $\zv$ is the power system measurements (without the coordinated FDI attack), as detailed in Section~\ref{sec:Preliminaries}. The notation $\yv$ are labels representing the location of the line under outage. The elements of the label $ y_l$ are given by
\begin{align}
    y_l = \Bigg \{ \begin{matrix} & 1, & \text{Line } \ l \  \text{is in outage} \\ & 0, & \text{otherwise,} \end{matrix} \label{eqn:y_rule}
\end{align}
where subscript $l \in L$ denotes the transmission line index. 

We note that multiple line outages can occur simultaneously. Hence, the number of potential line outage scenarios increases rapidly with the size of the power system. Enumerating all line outage scenarios will be computationally intractable. So, the traditional multi-classification method is not suitable for line outage identification. To overcome this difficulty, we recast the problem into $L$ binary classification problems, where each one corresponds to inferring the status of a line that is under outage separately. More specifically, let $\mathcal{T} = \{ \zv_{k}, \yv_{k} \}^{K}_{k = 1}$ denote the training dataset (input-output pairs). Herein, $K$ denotes the number of training samples and subscript $k$ denotes the training sample's index. Let $\hv(\zv_k,\wv)$ denote a parametric function, and $\wv$ denotes the parameters of the DNN. The output of the DNN can be represented as $\hat{\yv}_k =  \hv(\zv_k,\wv)$. The parameters of the DNN are trained to minimize the objective function given by
\begin{align}
    J_{\mathcal{T}} (\wv) = -\frac{1}{K} \sum_{k=1}^{K} \frac{1}{L} \sum_{l=1}^{L} & (\hat{y}_{l, k} log(y_{l, k}) \nonumber\\
     &+ (1-\hat{y}_{l, k })log(1-y_{l, k})).
    \label{eqn:loss_ML}
\end{align}
This objective function is the sum of $L$ binary cross-entropy between DNN's predicted value $\hat{\yv} =  \hv(\zv,\wv)$ and the corresponding ground truth labels $\yv$. 

\emph{Challenges of CCPA Localization Using DNN:} As opposed to the scenario described above, localizing the line disconnected under CCPA is challenging because the line outage is carefully masked by a coordinated FDIA. Our results show that the straightforward application of ML algorithms can only localize ``partial CCPA'' but can not localize ``extra CCPA'' and ``full CCPA''. The reason being that ``partial CCPA'' \eqref{eqn:CCPA_P} only removes the $ \Delta \Hm \bm{\theta}_p$ component from $\zv_p,$ while leaving the $\Hm \Delta \bm{\theta}$ component. When trained with the data corresponding to $\zv_{ccpa},$ DNN can build a mapping between $\Delta \bm{\theta}$ and the line that is disconnected by the attacker. However, the ``extra CCPA'' in  \eqref{eqn:CCPA_E} adds an additional component $\cv$ (i.e., $\Delta \bm{\theta} + \cv),$ and distorts the mapping between $\Delta \thetav$ and the line under outage. Finally, the ``full CCPA'' \eqref{eqn:CCPA_F} completely masks the physical attack vector, and the observed measurement $\zv$ corresponds to the normal system state (no attack), making it impossible to distinguish between the lines that are physically disconnected by the attacker. 

\subsection{Moving target defense} \label{sec:Moving target defense}

In the subsection, we propose strengthening the DNN-based line outage localization in the presence of a coordinated FDIA using MTD. The main idea is that to conduct a stealthy attack (partial, extra, and full CCPA), the attacker must obtain at least a partial knowledge of the target system topology (branch connectivity and line reactances). MTD is a dynamic defense strategy that changes the transmission line reactance periodically (or event-triggered) using D-FACTS devices to invalidate the attacker's acquired knowledge of the system (e.g., using data-driven methods \cite{LakshDataDrive2021}). The two essential steps in the construction of an MTD against CCPA are D-FACTS deployment and operation, which we describe in the following. 

{\bf D-FACTS deployment:} Reference \cite{Lakshminarayana2021} proposes an optimal deployment of D-FACTS devices to defend against CCPAs. Let us denote the set of transmission lines by $\mathcal{L}_D \subseteq \mathcal{L}$ on which D-FACTS devices are deployed. To launch an undetectable CCPA, the attacker requires the knowledge the branch reactances within at-least one alternative path between the two ends of disconnected lines. D-FACTS deployment should guarantee that at least one D-FACTS device is installed on each alternative path \cite{Lakshminarayana2021}. To find the minimum number of D-FACTS devices that can ensure this throughout the power grid, the problem can be solved by finding a spanning tree of the graph  $\gv = (\mathcal{N}, \mathcal{L})$, denoted by $\mathcal{L}_{sptr}$, and the D-FACTS deployment can be selected following $\mathcal{L}_D = \mathcal{L} \setminus \mathcal{L}_{sptr}$.

{\bf D-FACTS Operation:} MTD operates to change the transmission line reactance on $\mathcal{L}_D$, which represents as $\xv_{\mathcal{L}_D} = \{x_l \vert \forall l \in \mathcal{L}_D \}$, via D-FACTS devices.  We use the superscript `` $\prime$ '' to represent system notations after operating MTD (e.g., $\Hm$ and $\Hm^\prime$ denote topology matrix before and after MTD, respectively). The line reactance setting after MTD becomes $\xv^\prime_{\mathcal{L}_D} = \xv_{\mathcal{L}_D} + \Delta \xv_{\mathcal{L}_D}$, where $\Delta \xv_{\mathcal{L}_D}$ is the reactance perturbation. The effectiveness of MTD operation can be quantify using \textit{smallest principal angle} (SPA) between the column space of the topology matrix before and after MTD operations, i.e. $\gamma (\Hm, \Hm^\prime)$ \cite{LakshMTD2021}. MTD perturbations $\Delta \xv_{\mathcal{L}_D}$ that ensure a high value of $\gamma (\Hm, \Hm^\prime)$ are more effective in terms of the detection rate. 

Next, we investigate MTD in the context of the different CCPA scenarios described in the previous section. As in previous work on MTD, we assume that the attacker has an outdated knowledge of the system, i.e., system corresponding to the measurement matrix $\Hm$ (and correspondingly $\thetav,$ etc.). However, due to the MTD, the system's settings are changed to the measurement matrix corresponding to $\Hm^\prime$  (and correspondingly $\thetav^\prime,$ etc.). We exclude ``partial CCPA'' from our discussion since they can be directly localized by the DNN (without MTD, see discussion in the previous section).

For ``extra CCPA'' \eqref{eqn:CCPA_E}, the attacker computes the FDIA with outdated system knowledge. After MTD, the relationship between  $\zv'$ and $\zv'_p$ becomes  $\zv' = \zv'_p + \Hm' \Delta \bm{\theta}' + \Delta \Hm' \bm{\theta}_p'.$ The attacker in turn computes the ``extra CCPA'' as $\av_{extra} = \Delta \Hm  \bm{\theta}_p + \Hm \cv.$ The measurements following CCPA are given by 
\begin{align}
    \zv'_{ccpa\_extra} & = \zv'_p + \av_{extra} \nonumber \\
     & = \zv' + \Hm' \Delta \bm{\theta}' + \Delta \Hm' \bm{\theta}_p' - \Delta \Hm  \bm{\theta}_p + \Hm \cv \nonumber\\ 
    &= \zv' + (\Hm' \Delta \bm{\theta}' + \Hm \cv) + (\Delta \Hm' \bm{\theta}_p' - \Delta \Hm  \bm{\theta}_p) \label{eqn:CCPA_E_MTD}.
\end{align}
For ``full CCPA'' \eqref{eqn:CCPA_F}, to obtain the attack vector $\av_{full} =\zv -\zv_p$, the attacker should firstly recompute power flow equations, according to their knowledge of the target system. With MTD, the attackers' recomputed measurements are different from the real value, and thus the stealthy attack is invalidated, as follows:
\begin{align}
    \zv'_{ccpa\_full} &= \zv'_p + \av_{full} \nonumber\\
    & = \zv'_p + \zv - \zv_p. \label{eqn:CCPA_F_MTD}
\end{align}
In both cases, a DNN can be trained to learn the mapping from $\zv'_{ccpa\_extra}$ and $\zv'_{ccpa\_full}$ to the line under outage (i.e., disconnected by the attacker in a CCPA). We note that placement of CCPA as described above ensures that the knowledge of $\Delta \Hm^\prime$ is invalidated, and MTD chosen according to the SPA criteria ensures a high detection rate. We show the effectiveness of the proposed method in localizing CCPAs using simulations in Section \ref{sec:Simulation Results}.

\section{Meta Learning For Attack Localization Under MTD Topology Reconfigurations} \label{sec:Meta Learning}
We note that MTD-based topology reconfigurations will render an offline-trained DNN ineffective. E.g., when the topology of the system is changed from $\Hm$ to $\Hm^\prime,$ a DNN trained on parameters corresponding to $\Hm$ is no longer guaranteed to be effective. While it is certainly possible to retrain the DNN model when the topology reconfiguration to $\Hm^\prime$ is triggered, this approach requires significant amounts of training data and time. Alternatively, separate DNN models can be trained, as backup, for each system configuration. But this would require a significant amount of computational resources, and it is almost impossible to enlist all the system configurations beforehand. To overcome this challenge, we propose to apply meta-learning to assist the combination of DNN and MTD. 

Meta-learning is a DNN-training methodology for a series of related tasks; when presented with a new and related task, the strengthened model can quickly learn this task from a small amount of training data samples \cite{FinnMAML2017}. In particular, the CCPA localization under the different system topologies (due to MTD topology reconfiguration) can be viewed as a series of related learning tasks. Meta-learning enables DNN to achieve fast retraining and quickly adapt to new system configurations. Specifically, we use the so-called model agnostic meta-learning (MAML) \cite{FinnMAML2017}, which finds a common initialization point in the offline training phase for a series of related tasks (in our context line outage localization under different topologies). Then, in the online training phase, following a topology reconfiguration, a DNN model can quickly learn the new task with only a few training samples by starting with this initial point. 

 \begin{figure}[!t]
	\centering
	\includegraphics[width=0.4\textwidth]{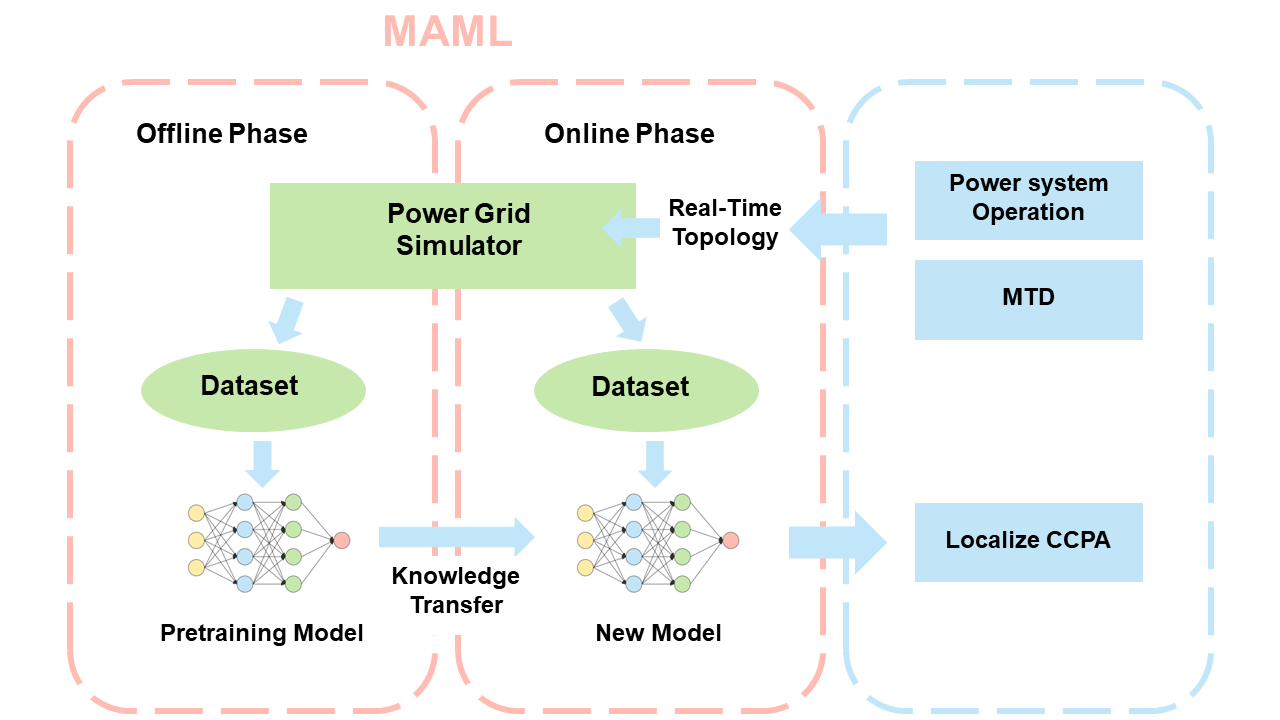}
	\caption{Overall structure of CCPA localization}
	\label{fig:Coordinated DNN-MTD}
\end{figure}

The overall structure of the CCPA localization model with meta-learning is shown in Figure~\ref{fig:Coordinated DNN-MTD}. In the offline training phase, the system operator uses a power grid simulator to generate a large amount of line outage localization training data under different system topologies (e.g., these topologies can be generated by adding random reactance perturbations to the original system model). The data is subsequently used to develop a pretraining model using the MAML algorithm. When MTD operates to perturb the transmission line reactance and modify the system topology, system operators will use a power grid simulator to generate a few data samples from the real-time system topology (note these data samples can be computed using a simulator once the reactance perturbation is computed by the system operator). The real-time samples can be used to quickly fine-tune the pretraining model. The developed new model can identify CCPA location under the real-time system.  We omit the detailed description of the algorithm due to the lack of space and refer the reader to our past work \cite{Chen2022}, in which we apply meta-learning for a similar regression task in power systems. 

\section{Simulations } \label{sec:Simulations Settings}

In this section, we perform simulations to verify the effectiveness of the proposed approach in identifying the location of physical attacks in CCPA. 

\subsection{Simulation Settings} \label{sec:Simulations Settings A}
The proposed CCPA identification approach is tested using the IEEE-14 bus system. The power system model is obtained from the MATPOWER simulator. Following the optimal D-FACTS deployment algorithm to defend against CCPAs \cite{Lakshminarayana2021}, D-FACTS devices are placed on 7 links given by \{1,3,5,8,9,18,19\} and the reactance perturbation for MTD is chosen to obtain a large SPA between the pre-perturbation and the post-perturbation matrices. 

The compromised measurement data is generated by injecting attack vectors into the original measurement data. We generate the aforementioned ``partial CCPA'', ``extra CCPA'', and ``full CCPA''. To create measurements under CCPA, we firstly disconnect a randomly-chosen subset of transmission lines. We assume that at-most two lines can be under outage (note that if too many lines are disconnected, it might lead to an infeasible power flow solution). After that, compromised measurement data is generated by injecting attack vector \eqref{eqn:CCPA_P}, \eqref{eqn:CCPA_E}, \eqref{eqn:CCPA_F} in the absence of MTD and according to \eqref{eqn:CCPA_E_MTD} and \eqref{eqn:CCPA_F_MTD} when MTD is implemented.  To generate multiple data points, we simulate the system under different load conditions by randomly changing the active load within $[0.8, 1.2]$ of the base values (here, the base value refers to the value provided in MATPOWER case file). For each scenario, we create 10000 training samples and 1000 testing samples.

The pre-training algorithm of meta-learning requires data from different system topologies. For this, we inject random perturbation to the line reactance values ranging between  $[0.8, 1.2]$ of their base values (once again, base values refer to the reactance value provided in MATLAB case file). Specifically, we generate $100$ different topologies. For each topology, we generate $1000$ data points by changing the load conditions. For real-time retraining following MTD, a reactance perturbation is chosen according to the procedure described above, and the DNN is trained online starting from the initial point specified by meta-learning.  

We implement the DNN model using PyTorch. Specifically, a multi-layer convolution neural network (CNN) is designed to process the data. The CNN setting is detailed in Table~\ref{tbl:1-D CNN}. We use the ReLu activation function at the hidden layers and the sigmoid activation function at the output layer. The model parameters are updated using the ``Adam'' optimizer, and the L2 regularization is applied to prevent over-fitting. 

\begin{table}[!t]
 	\centering
     	\begin{tabular}{|c|c|c|c|c|c|}
 		\hline
 		Layer & Type & Input size & Output size & Kernel & Padding \\
 		\hline
 		1 & Conv1d & $ 1 * 54 $ & $ 128 * 54 $ & 5*1 & 2 \\
 		\hline
		2 & Conv1d & $ 128 * 54 $ & $ 256 * 54 $ & 3*1 & 1 \\
		\hline
 		3 & Conv1d & $ 256 * 54 $ & $ 128 * 54 $ & 3*1 & 1 \\
 		\hline
 		4 & Conv1d & $ 128 * 54 $ & $ 64 * 54 $ & 3*1 & 1 \\
		\hline
 		5 & Linear  & 64 * 54 & 20 & & \\
 		\hline
 	\end{tabular}
 	\caption{The CNN setting}
 	\label{tbl:1-D CNN}
 \end{table}

\subsection{Simulation Results} \label{sec:Simulation Results}
The performance of developed models is assessed using precision rate and recall rate, which are defined as
\begin{align}
    precision = {{True\ Positive} \over {True\ Positive + False\ Positive}},
\end{align}
\begin{align}
    recall = {{True\ Positive} \over {True\ Positive + False\ Negative}}.
\end{align}

The overall identification performances of the developed models are presented in Table~\ref{tbl:Result}. We compare the performance of three approaches in identifying line outage locations in CCPAs. Approach~1 is the straightforward application of ML algorithms (i.e. CNN) using the compromised measurements to identify the line outage locations. When defending against ``partial CCPA'', Approach~1 achieves accurate identification performance with $ 96.09 \% $ recall rate and $ 99.23\% $ precision rate. However, when attackers intentionally inject noise into $\Delta \thetav$ (i.e. ``extra CCPA''), the recall rate rapidly decreased to $ 64.70 \%$. When it comes to ``full CCPA'', Approach~1 can hardly identify the attack.

In Approach~2, MTD is used to invalidate the cyber mask. Following MTD reconfiguration, we retrain the CNN from scratch (i.e., with randomly initialized weights) and with the data corresponding to the new system configuration. The effectiveness of MTD can be validated by comparing the performance of Approach~2 and Approach~1, especially in the case of ``extra CCPA'' and ``full CCPA''. We note that the recall rate exceeds $83 \%$, and the precision rate exceeds $95 \%$. These results prove that MTD improves the attack localization performance and contributes to the defense against stealthy CCPA.

However, CCPAs with knowledge distortion (due to MTD) still result in a disturbance in attack localization, which is equivalent to topology identification from noisy measurement. Our proposed approach (Approach~3) applies MAML to strengthen the CNN and help denoising. Compared with Approach~2, our approach achieves up to $7 \%$ and $3 \%$ improvement in recall rate and precision rate, respectively. Furthermore, our approach increases the CNN training speed and reduces the data requirements, which enables the model to adapt to topology reconfiguration caused by MTD.

\newcommand{\tabincell}[2]{\begin{tabular}{@{}#1@{}}#2\end{tabular}}

\begin{table}[!t]
	\centering
    	\begin{tabular}{|c|c|c|c|}
		\hline
		Approach & Attack & Recall Rate (\%) & Precision Rate (\%) \\
		\hline
		\multirow{3}*{CNN} & P & 96.09 & 99.23 \\
		\cline{2-2}
		 & E & 64.70 & 92.21 \\
		\cline{2-2}
		 & F & 3.30 & 10.40\\
		\hline
	    \multirow{3}*{CNN+MTD} & P & 95.80 & 99.86\\
		\cline{2-2}
		& E & 83.09 & 95.52 \\
		\cline{2-2}
		& F & 83.72 & 92.61  \\
		\hline
		\multirow{3}*{\tabincell{c}{CNN+MAML\\+MTD}} & P & 97.16 & 99.55 \\
		\cline{2-2}
		& E & 87.85 & 96.20 \\
		\cline{2-2}
		& F & 91.31 & 95.86 \\
		\hline
	\end{tabular}
	\caption{Simulation results for IEEE-14 bus system. ``P'' denotes ``partial CCPA'', ``E'' denotes ``extra CCPA'', ``F'' denotes ``full CCPA''.}
	\label{tbl:Result}
	\vspace{-0.1 cm}
\end{table}

\section{Conclusions} \label{sec:Conclusions}

In this work, we have proposed a novel approach to localize the line(s) disconnected by the attacker in CCPA. The proposed approach is a combination of data-driven defense and physics-based defense. Specifically, MTD is applied to invalidate the knowledge the attackers use to mask the effect of the physical attack and MAML-strengthened CNN is used to localize the line outage. Extensive simulation results verify the effectiveness of the proposed approach in defending against stealthy CCPAs. To the best of our knowledge, this work is the first to localize the disconnected lines in CCPA. Future work includes deriving theoretical results for optimal MTD design for localization purposes and extension to a non-linear power flow model.  

\bibliographystyle{IEEEtran}
\bibliography{IEEEabrv,bibliography}

\end{document}

%% file: macros.tex
\newfont{\bbb}{msbm10 scaled 500}

\newfont{\bb}{msbm10 scaled 1100}

\newcommand{\RR}{\mbox{\bb R}}

\newcommand{\av}{{\bf a}}

\newcommand{\cv}{{\bf c}}

\newcommand{\ev}{{\bf e}}

\newcommand{\gv}{{\bf g}}
\newcommand{\hv}{{\bf h}}

\newcommand{\rv}{{\bf r}}

\newcommand{\wv}{{\bf w}}

\newcommand{\xv}{{\bf x}}
\newcommand{\yv}{{\bf y}}
\newcommand{\zv}{{\bf z}}

\newcommand{\Am}{{\bf A}}

\newcommand{\Dm}{{\bf D}}

\newcommand{\Fm}{{\bf F}}

\newcommand{\Hm}{{\bf H}}

\newcommand{\Pm}{{\bf P}}

\newcommand{\Wm}{{\bf W}}

\newcommand{\thetav}{\hbox{\boldmath$\theta$}}

\newcommand{\diag}{{\hbox{diag}}}

\newcommand{\defines}{{\,\,\stackrel{\scriptscriptstyle \bigtriangleup}{=}\,\,}}